\documentclass{IEEEtran}
\usepackage{cite}
\usepackage[pdftex]{graphicx}
\usepackage{amsmath}
\usepackage{amsfonts}
\usepackage{algorithmic}
\usepackage{algorithm}
\usepackage{array}
\usepackage{dblfloatfix}
\usepackage{url}
\usepackage{subfigure}
\usepackage{flushend}

\newcommand\MYhyperrefoptions{bookmarks=true,bookmarksnumbered=true,
pdfpagemode={UseOutlines},plainpages=false,pdfpagelabels=true,
colorlinks=true,linkcolor={blue},citecolor={blue},urlcolor={blue},
pdftitle={Analysis of a heterogeneous social network of humans and cultural objects},
pdfsubject={Online Social Network Analysis},
pdfauthor={S. Agreste, P. De Meo, E. Ferrara, S. Piccolo, A. Provetti},
pdfkeywords={aNobii, Online Social Network Analysis}}
\usepackage[\MYhyperrefoptions,pdftex]{hyperref}
\usepackage{tikz}
\usepackage{setspace}
\def\R{{\cal R}}

\long\def\HIDE#1\ENDHIDE{\message{(Commented text...)}\par}
\newtheorem{definition}{Definition}

\begin{document}
\title{Analysis of a heterogeneous social network of humans and cultural objects}

\author{Santa~Agreste,
        Pasquale~De~Meo,
        Emilio~Ferrara,$^*$
        Sebastiano~Piccolo,
        and~Alessandro~Provetti
\IEEEcompsocitemizethanks{\IEEEcompsocthanksitem S. Agreste and A. Provetti are with the Dept. of Mathematics and Informatics, University of Messina. I-98166 Messina, Italy. 
E-mail: \{sagreste,ale\}@unime.it%
\IEEEcompsocthanksitem P. De Meo is with the Dept. of Ancient and Modern Civilizations, University of Messina. I-98166 Messina, Italy. 
E-mail: pdemeo@unime.it
\IEEEcompsocthanksitem E. Ferrara is with the Center for Complex Networks and Systems Research at Indiana University, Bloomington IN 47408 USA. 
$^*$Corresponding author. E-mail: ferrarae@indiana.edu}
\thanks{Manuscript received July 26, 2013; revised February 4, 2014.}}
\markboth{Transactions on Systems, Man and Cybernetics: Systems,~Vol.~N, No.~NN, Month~2014}%
{Agreste \MakeLowercase{\textit{et al.}}: Analysis of a heterogeneous social network}

\IEEEtitleabstractindextext{%
\begin{abstract}

Modern online social platforms enable their members to be involved in a broad range of activities like getting friends, joining groups, posting/commenting resources and so on.
In this paper we investigate whether a correlation emerges across the different activities a user can take part in.
To perform our analysis we focused on aNobii, a social platform with a world-wide user base of book readers, who like to post their readings, give ratings, review books and discuss them with friends and fellow readers.
aNobii presents a heterogeneous structure: i) part social network, with user-to-user interactions, ii) part interest network, with the management of book collections, and iii) part folksonomy, with books that are tagged by the users.
We analyzed a complete and anonymized snapshot of aNobii and we focused on three specific activities a user can perform, namely her tagging behavior, her tendency to join groups and her aptitude to compile a wishlist reporting the books she is planning to read.
In this way each user is associated with a tag-based, a group-based and a wishlist-based profile.
Experimental analysis carried out by means of Information Theory tools like entropy and mutual information suggests that tag-based and group-based profiles are in general more informative than wishlist-based ones.
Furthermore, we discover that the degree of correlation between the three profiles associated with the same user tend to be small.
Hence, user profiling cannot be reduced to considering just any one type of user activity (although important) but it is crucial to incorporate multiple dimensions to effectively describe users preferences and behavior.

\end{abstract}

}

\maketitle

\IEEEdisplaynontitleabstractindextext

\section{Introduction}\label{sec:intro}

Online social platforms are widely acknowledged as privileged venues where users can generate, share and consume content.
The ability of collecting large amount of data describing user activities in such platforms offers unprecedented opportunities to investigate whether different user behaviors are somewhat related \cite{santos2009individual,Aiello*10,aiello2012people,Aiello*13,demeo2013analyzing}.
Understanding the interplay between user activities is still an open research problem with important implications, including building better recommender systems  \cite{zhou2010solving,DeMeo*11,ma2011recommender,kazienko2011multidimensional}, predicting patterns of collective attention, behavior and communication \cite{centola2010spread,lerman2010information,weng2012competition,conover2013geospatial} or unveiling the dynamics of group formation and evolution in technologically-mediated social networks \cite{ferrara2011topological,ferrara2012large,grabowicz2012social,conover2013digital}.

Many authors \cite{mcpherson2001birds,aiello2012friendship} highlighted an alignment between social structures generated from user interactions and attributes describing user features and tastes.
Such an alignment broadly falls in the scope of the {\em homophily theory}, which suggests that people are more likely to create social connections with others sharing their interests.

The social structure created by people with similar interests is a powerful tool to spread ideas and behavior: recent studies \cite{bollen2011happiness,golder2011diurnal} investigated how happiness propagates over the members of Twitter and they show that such a propagation is regulated by homophily.
Analyzing the correlation between social links and personal mood may become a key step to understand the collective mood of large populations, which, in turn, plays a major role in influencing individuals' behavior and collective decision making.

However, uncovering the relationship binding social structures and user interests is a hard task for a number of reasons.
First of all, information describing user contributed contents ({\em semantic signals}) as well as social ties ({\em social signals}) should be available.
Semantics and social signals can be acquired by crawling an online social platform \cite{ferrara2011crawling,gjoka2011practical} but the reliability of the findings from the analysis of a collected sample crucially depends on its size and significance.
The effects generated by wrongly collecting data may be disruptive \cite{handcock2010modeling,morstatter2013sample}.

A further drawback is that, in many cases, user feedback can be conceived as a combination of social and semantic signals whose separation is hard and they can be easily confounded \cite{shalizi2011homophily}.
For instance, consider the comments users are posting about an object they recently bought.
Comments have a semantic connotation because their linguistic analysis highlights how a particular person reacts to a given topic she is exposed to \cite{danescu2013no}.
However, from a homophily perspective, users may avail themselves of the comments posted by others to be kept abreast of novel content circulating in their social sphere and to quickly form an opinion about unknown items.

In the past, few authors studied the relationship between social and semantic signals \cite{santos2009individual,schifanella2010folks}.
Santos \emph{et al.} \cite{santos2009individual} focused on {\sl CiteULike} and {\sl Connotea,} two social sharing systems aiming to promote the sharing of scientific publications.
The authors provided a measure of user similarity based on user tagging activities and they studied whether user similarity relates to social behavior (\emph{e.g.,} the level of engagement in a discussion group.)
That work, however, focused on a specific class of Web users (\emph{i.e.,} researchers) which are strongly motivated to access bibliographic repositories for professional reasons.
A different context was explored by Schifanella \emph{et al.} \cite{schifanella2010folks} who focused on two popular online platforms, \emph{i.e.,} {\sl Last.Fm} and {\sl Flickr} and studied the possibility to predict social links between two users on the basis of the similarity of their tags.

Currently, very little is known about the correlation among user activities when multiple signals are available and some of these signals present both a social and a semantic component.
This article attempts at filling this knowledge gap: to this purpose, we extensively analyzed {\em aNobii}, a Social Web platform hosting book lovers.
\emph{aNobii} is, in our opinion, an ideal test-bed: first of all, a {\em complete} and anonymized snapshot of the aNobii social network has been collected by Aiello \emph{et al.} \cite{Aiello*10} and it was made available for research purposes.
Hence, we can study the whole aNobii social network rather than a sample extracted from it.
In addition, aNobii user activity provides both social and semantic signals, albeit in combination.
aNobii users may join (or start) {\em thematic groups}, as well as create friendships with other users.
They are also allowed to review books and label them by means of tags, either freely-chosen or from a pre-defined set of categories.
Finally, a nice feature of aNobii is the presence of {\em wishlists}, \emph{i.e.,} a collection of books a user is planning to read.

Wishlists are halfway through semantic and social signals: on the one hand, a wishlist specifies the books a user is interested to.
On the other, wishlists represent a socially-inspired tool for helping users to find books of their interest.
In fact, reading can have a social aspect and, in some cases, pleasant readings are those suggested by our peers or friends.
Users can browse each others wishlists and discover new books, read their reviews, comments and ratings; users are also allowed to exchange/donate/sell books.
In this way, they can get valuable information even if they are not able to clearly formulate their needs; this feature increases the chances of serendipitous finding of books of their interest without even explicitly querying the platform.

The results of our analysis may be useful in a large number of applications.
First of all, we could use the results of our analysis to understand how users perceive a specific platform: in case of aNobii, for instance, if relevant discrepancies would emerge in the intensity of tagging activity with respect to social activities, we may conclude that users would mainly perceive aNobii as a tool to better organize and manage their private book collection.
By contrast, if the intensity of social activities would dominate over tagging ones, we could think aNobii as a platform capable of fostering social aggregation and collaboration among users.
The output of our analysis is also useful to develop novel and unconventional strategies to recommend items in a social platform.
More concretely, if we are interested in predicting the books that will appear in the bookshelf of a given user, we could establish in advance if it is more advantageous to look at the tags she applied, the groups she joined or the books in her wishlist (or combination of these dimensions.)

To perform our analysis we associated each user with three profiles, namely: {\em (i)} a {\em tag-based profile} (describing her tagging behavior), {\em (ii)} a {\em group-based profile} (specifying her tendency to join groups) and {\em (iii)} a {\em wishlist-based profile} (containing information about her wishlist.)
Each of these profiles specifies a dimension of analysis.

\smallskip
\noindent
The main findings of our analysis are the following:

\begin{enumerate}

\item In all the dimensions under investigation we found a small fraction of users which were heavily involved in activities like tagging, joining groups and compiling wishlists. By contrast, the large amount of remaining users was mostly inactive and their profiles were quite sparse. We also found that the average size of user profiles was not uniform across each of the dimensions and, in particular, the highest level of variability in the size of user profiles occurred in wishlist-based profiles.

\item We studied the correlation among user profiles in different dimensions by computing the Spearman's $\rho$ coefficient. The resulting $\rho$ was in general quite small, independently of the dimensions under investigation. Such an observation has practical consequences: in fact, if we wish to provide personalized services, it would be crucial to integrate information coming from different profiles into a global one in order to get a better picture of user needs.

\item We studied the information content embodied in each user profile by computing its Shannon's {\em information entropy}. Our analysis revealed that tag-based and group-based profiles were, in general, more informative than wishlist-based ones. We also took user profiles in pairs and we computed their mutual information: we observed that the profiles share little mutual information, which means that the knowledge of the profile of a user in only one dimension is not sufficient to predict her behavior in other dimensions.

\item Finally, we observed a small but non-negligible fraction of users showing extreme behavior, \emph{i.e.,} users who are very prolific in a particular dimension but who were almost inactive in the others. Extreme behavior occurred in the practice of tagging and this informs us that a portion of users perceives aNobii as a tool for organizing knowledge rather than for creating social contacts.

\end{enumerate}

%


\section{Related Work}\label{sec:related}
In the latest years, research on online social networks (hereafter, OSNs) and social tagging received a lot of attention.
A growing number of online social platforms (among which we include aNobii) allows their members to be involved in a wide range of activities like creating social links, joining groups, tagging/commenting/rating objects.
These platforms are often modeled as {\em heterogeneous social networks} (HSNs) (also known as {\em multidimensional social networks}, {\em multi-layered} or {\em multi-relational social networks}) \cite{kazienko2011multidimensional,kazienko2010multi,tang2012inferring,berlingerio2013multidimensional}.

The heterogeneity of HSNs mainly depends on two factors (or combinations of them.)
On the one hand, several social relationships can link two members of a HSN: for instance, two users can be connected because they are friends and, {\em simultaneously}, colleagues. On the other, multiple relationships between two users can be inferred by monitoring their behaviors: for example, two users could be regarded as connected because they joined the same group. For each activity we can thus extract a social network and, then, a HSN appears as a juxtaposition of multiple networks, each recording a specific kind of user activity.

Some researchers focused on the analysis of the structural properties of a HSN and studied the interplay of user behavior in a HSN.
Other authors were more concerned on predicting new social links or recommending items in a HSN. Here we review related literature in the domain of HSNs and aNobii.

\subsection{Analysis of Heterogeneous Social Networks}
\label{sub:analysis-HSN}

Recently, some authors \cite{santos2009individual,schifanella2010folks} started studying the interplay of social activities and user-generated semantics.
In \cite{santos2009individual}, the authors focused on two datasets, {\em CiteULike} and {\em Connotea} about scientific publications.
In order to quantify the strength of social ties involving users, they defined the {\em level of collaboration} of two users as the number of discussion groups to which two users were jointly affiliated.
In an analogous fashion, the similarity between a pair of users was defined as the number of tags they jointly exploited or the number of scientific articles read by both of them.
That work shows that the {\em lack of collaboration} was related to the {\em lack of joint interests}: users who never read/bookmarked the same group of papers also tend to join different discussion groups.

A further analysis is provided in \cite{schifanella2010folks}.
In that paper, the authors conducted extensive experiments on data samples extracted from Last.Fm and Flickr.
They showed that a strong correlation exists between user activity in a given social context (\emph{e.g.,} the number of friends of a user or the number of groups she joined) and the tagging activity of the same user.
The most active users tend to have as neighbors users who are also highly active.
As a final result, the authors observed that an alignment between users' tag vocabularies emerges between close users in the social network.

Our paper advances the state of the art in the analysis of HSNs in a number of directions.
First of all, we consider semantic behavior (encoded by user tagging behavior), social behavior (considering the tendency of users to join the same groups) and user behavior that provides both a social and a semantic component (\emph{i.e.,} the wishlist users expose.)
We study not only the correlation between different users' behavior but also the information content associated with each profile by means of Information Theory tools like entropy and mutual information.


\subsection{Predicting Social Links and Recommending Items in HSN}
\label{sub:link-prediction-recommendation-hsn}

The task of predicting and recommending new social links as well as items in a HSN has attracted the interest of many researchers \cite{DeMeo*11,wang2011learning,diehl2007relationship,wang2010mining,tang2012inferring,chen2008combinational}.
Some authors \cite{DeMeo*11,wang2011learning} suggest to combine multiple data sources describing user activities in a social platform to better quantify how a given item can be relevant to a particular user or to predict new social links.

A nice approach to recommend social links has been discussed by Kazienko \emph{et al.} \cite{kazienko2011multidimensional}.
The authors studied Flickr and identified 11 types of relations binding users (\emph{e.g.,} relations based on contact lists, relations based on tags, relations based on opinions about photos and so on.)
Each relation $r_k$ is used to compute the strength $s_{ij}^{k}$ of the social tie between two users $u_i$ and $u_j$; for a fixed relation $r_k$, user community can be represented as an undirected and weighted graph such that each vertex represents a user $u_i$ (resp., $u_j$) and the weight of the edge connecting $u_i$ and $u_j$ equals $s_{ij}^{k}$.
Strength values computed in each layer are aggregated to compute the global strength value and a further weight is introduced to compute the relevance that a given layer has for $u_i$.
These weights are then used to calculate the value $v_{ij}$ that $u_i$ would gain if $u_j$ were recommend to her.

Other authors focused on the task of determining the meaning of a social relationship and predicting new social links.
For instance, in \cite{diehl2007relationship} the authors provide a supervised ranking approach to identifying relationship in a network consisting of the employees of a company with the goal of discovering subordinate-manager relationships.
Another approach \cite{wang2010mining} analyzes a co-authorship network in the Computer Science domain and provides a probabilistic model to discover the roles of authors as well as the advisor-advisee relationships.

The strategies described above mainly focus on a specific domain.
A more general approach is discussed in \cite{tang2012inferring}.
In that paper, the authors model a HSN as a graph $G$; an edge in $G$ can be labeled with multiple attributes like \emph{family}, \emph{colleague} or \emph{classmate}.
Some labeled relationships are available as training data and this information is instrumental to learn a predictive function that, for each edge $e$ in $G$, computes the probability that $e$ has a particular label.

Some researchers focused on the problem of recommending groups a user can join and, in such a case, a social link connects a user with a group she joined.
For instance, Chen {\em et al.} \cite{chen2008combinational} provide an algorithm called CCF (Combinational Collaborative Filtering) which is able to suggest new friendship relationships as well as the communities they could join.
CCF considers a community from two different but related perspectives: a community is a {\em bag of users} (formed by its members) and a {\em bag of words} describing community interests.
By merging different types of information sources it is possible to alleviate the data sparsity arising when only information about users (resp., on words) is used.

In \cite{chen2009collaborative} the authors propose two algorithms to recommend communities.
The former is based on association rule mining and it aims at discovering frequently co-occurring sets of communities.
The latter is based on {\em Latent Dirichlet Allocation} \cite{BlNgJo03}.
It aims at modeling user-communities co-occurrences by means of a latent model and use this model to predict the communities a user should join.

A further research stream aims at recommending items in a HSN.
For instance, the approach of Wang \emph{et al.} \cite{wang2011learning} focuses on the problem of predicting click-through rate (CTR) of a commercial advertisement (ad) for a given user or a given query.
In that paper, the text associated with the Web pages visited by the users is analyzed and topics are extracted by means of a classifier.
To train such a classifier, a set of labeled Web pages plus a reference ontology is used.
Extracted topics play the role of {\em concepts} and both user interests and the content of an ad are modeled as a vector of concepts: each user $\mathbf{u}$ and an ad $\mathbf{a}$ are associated with two vectors $\mathbf{v}_u$ and $\mathbf{v}_a$ respectively; here, the $i$-th component of $\mathbf{v}_u$ and $\mathbf{v}_a$ specify how much the $i$-th concept is relevant to $\mathbf{u}$ and $\mathbf{a}$.
The degree of matching $\sigma(\mathbf{u}, \mathbf{a})$ between $\mathbf{u}$ and $\mathbf{a}$ is defined as the inner product of $\mathbf{v}_u$ and $\mathbf{v}_a$.
The $\sigma(\mathbf{u}, \mathbf{a})$ coefficients are then used in conjunction with logistic regression to predict the probability that user $\mathbf{u}$ will click on ad $\mathbf{a}$.
In \cite{DeMeo*11} an unsupervised approach to recommend resources in HSN is proposed.
The authors model a HSN as a hypergraph whose vertices represent users and resources available in an OSN; hyperedges identify connections among users on the basis of their past activities.
A variant of the Katz coefficient \cite{katz1953new} is employed to determine the similarity degree of two users and, finally, to suggest both new social links and resources.
This approach also works in a {\em cross-social network} scenario, \emph{i.e.,} it assumes that users created multiple accounts on different online social platforms.
Finally, Chen \emph{et al.} \cite{ChZeYu13} propose a framework supporting different type of recommendations (item, friend and group recommendation.)
In this approach two type of information were combined to produce recommendations of the other type (\emph{e.g.,} information about friendship and items were merged to suggest groups.)
A matrix factorization algorithm was used to perform such a combination.


\subsection{Studies on aNobii}
\label{sub:studies-anobii}

The first extensive study of the aNobii platform and its features is due to Aiello {\em et al.} \cite{Aiello*10,aiello2012people,Aiello*13}.
In those papers, the authors studied the dynamics of link creation and the social influence phenomenon that may trigger the diffusion of information in aNobii.

Some of the findings of \cite{Aiello*10,aiello2012people,Aiello*13} are extremely interesting: for instance, the selection of social partners is strongly driven by structural, geographical, and topical proximity and this provides a good basis for developing an algorithm for recommending social links.

The analysis from Aiello \emph{et al.} differs significantly from that we carried out in this work because:
{\em (i)} in \cite{Aiello*10,Aiello*13} the authors considered as social links the friendship and neighborhood relationships, whereas here we focus on group affiliation.
We believe that group affiliation, besides highlighting social interactions among users, closely reflects user interests.
{\em (ii)} The goal of those papers \cite{Aiello*10,aiello2012people,Aiello*13} was to quantitatively analyze how book adoption spread across the aNobii social network therefore the authors tracked patterns of diffusion of specific information.
By contrast, our goal is to study the alignment between tag-based, group-based and wishlist-based profiles.

\section{The aNobii Sharing platform}\label{sec:preliminaries}
{\em aNobii}\footnote{{\em aNobii} official website: {\tt www.anobii.com}} is a social networking Web platform for book lovers and readers.
It was created in Hong Kong in 2005 and now its members are distributed across several countries.

aNobii users are allowed to provide information about their bookshelf as well as personal information like their gender, age, country and, optionally, their town.
According to Aiello {\em et al.} \cite{Aiello*13}, country is declared in about 97\% of user profiles; in addition, roughly 40\% of the profiles contains also an indication of town.
The analysis of geographical information reveals that the aNobii social network fragments in two main communities, namely Italy and countries in Asia (namely Taiwan, Hong Kong and China.)
About 60\% of user population resides in Italy while the percentage of aNobii users residing in the Asian countries is about 30\% of the whole aNobii population.
Connections between the two geographical communities are, however, quite sparse and they are mediated by smaller intermediate geographical clusters (\emph{e.g.,} the community of the users residing in the USA.)

The book collection mainly consists of two components: the list of books read by a given user and a {\em wishlist} containing the book titles she is planning to read.
Users can rate and review books as well as tag them.
Digital bookshelves resemble ``private libraries'' and, often, they are not specialized in a specific literary genre: in most cases, in fact, digital bookshelves contain popular novels plus essays, scientific books or comics, even if novel seems to be the prevailing literary genre.
In some cases users appear prone to include books written in foreign languages and, therefore, the linguistic borders that traditionally assign a book to a national literature appear therefore fuzzier.
This opens up to a {\em globalized} vision of literature and it is not surprising that books like the {\em Harry Potter} series or {\em The Lord of the Ring} trilogy are worldwide very popular.
Users are allowed to specify their {\em reading status} (\emph{e.g.,} a user may specify if she has finished to read a book.)

aNobii provides also some social features. Two types of user relationships are allowed: users can be {\em friends} if they know each other (for example, in real life or in other online social networks) or {\em neighbors} if the first one considers the library of the second one as relevant (this recalls the follower/followee relation in Twitter.)
The two relationships above are {\em mutually exclusive} and the creation of a link binding two users does not require their mutual approval (differently, for example, from Facebook.)
Therefore, links between two users can be regarded as {\em direct}.
If two users are linked (either as friends or neighbors), then the updates in the library of one of them are immediately notified to the other.

A further social option offered by aNobii is the possibility of forming and joining {\em thematic groups} (in short, {\em groups}.)
Groups are mainly used as discussion venues for book lovers of similar genres or for those who share similar reading interests.
Information about the groups a user joined, the books she read, and her social links are public.

\subsection{Notations and definitions}\label{sub:notations}
Here we introduce the notation that will be adopted throughout the remainder of the paper.
Let {\em (i)} $U$ be the set of aNobii users, {\em (ii)} $B$ be the set of available books, {\em (iii)} $T$ be the set of user-contributed tags, {\em (iv)} $G$ be the set of available groups and {\em (v)} $W$ be the set of available wishlists.
The sets $U$, $B$, $T$ $G$ and $W$ are henceforth called {\em dimensions}.

We are interested in modeling the strength of the relationship bonding two users according to a specific dimension.
For instance, if we focus on the $T$ dimension, we may assume that the larger the number of tags two users jointly applied, the stronger the relationship binding them according to the $T$ dimension.
In this paper we shall focus on three dimensions, namely $T$, $G$ and $W$.
We shall use {\em undirected and weighted} graphs to model relationship strength according to $T$, $G$ and $W$ dimensions and these three graphs will be denoted as $\mathcal{G}_{UT}$, $\mathcal{G}_{UG}$ and $\mathcal{G}_{UW}$.
In each of these graphs, each user $u_i$ is uniquely represented by a vertex $v_i$.
Edges are defined as follows:

\begin{enumerate}
   \item in $\mathcal{G}_{UT}$, $v_i$ and $v_j$ are linked if $u_i$ and $u_j$ jointly applied at least one tag.
   The weight $w_{UT}(v_i,v_j)$ of the edge $v_i$ and $v_j$ equals the number of tags that $u_i$ and $u_j$ jointly applied.

   \item in $\mathcal{G}_{UG}$, $v_i$ and $v_j$ are linked if $u_i$ and $u_j$ have at least one group in common.
   The weight $w_{UG}(v_i,v_j)$ of the edge joining $v_i$ and $v_j$ specifies the number of common groups $u_i$ and $u_j$ are jointly affiliated with.

   \item in $\mathcal{G}_{UW}$, $v_i$ and $v_j$ are linked if there is at least one book appearing in both the wishlist of $u_i$ and in the wishlist of $u_j$.
       The weight $w_{UW}(v_i,v_j)$ of the edge joining $v_i$ and $v_j$ indicates the overall number of books in the wishlist of $u_i$ that appear also in the wishlist of $u_j$.
\end{enumerate}

We are now able to introduce of the concept of {\em user profile on a specific dimension}:

\smallskip

\begin{definition}\label{def:userprofile}
Let $u_i \in U$ be a user.
The {\em tag-based} $\vec{p}_T(i)$, the {\em group-based} $\vec{p}_G(i)$ and the {\em wishlist-based profile} $\vec{p}_W(i)$ of $u_i$ are vectors in $\R^{|U|}$ and they are defined as follows

\begin{equation}\label{eqn:user-prof-def}
\begin{split}
\vec{p}_T(i) = & \{\vec{p}_T(i) \in \R^{|U|}: \vec{p}_T[i,j] = w_{UT}(u_i,u_j)\}\\
\vec{p}_G(i) = & \{\vec{p}_G(i) \in \R^{|U|}: \vec{p}_G[i,j] = w_{UG}(u_i,u_j)\}\\
\vec{p}_W(i) = & \{\vec{p}_W(i) \in \R^{|U|}: \vec{p}_W[i,j] = w_{UW}(u_i,u_j)\}\\
\end{split}
\end{equation}

\end{definition}

\smallskip

In our framework, the profile of a user is a {\em multidimensional entity} because we associate each user with as many profiles as the dimensions we are willing to observe.

\subsection{Description of the Dataset}\label{sub:dataset}
Our analysis was carried out on a complete snapshot of aNobii collected by Aiello \emph{et al.} \cite{Aiello*10} in September 2009.
The main features of available data are reported in Table \ref{tab:main-anobii}.

{\small
\begin{table}[htpb]
\centering
\caption{Features of the aNobii dataset} \label{tab:main-anobii}
\begin{tabular}{|c|c|}
\hline
\hline
Number of Users & 81,218\\
\hline
Number of Users who applied & 25,060\\
at least one tag & \\
\hline
Number of Users who joined & 41,833\\
at least one group & \\
\hline
Number of Users who applied at least & 18,039\\
one tag and joined at least one group & \\
\hline
Number of Books & 1,548,511\\
\hline
Number of Tags & 5,106,207\\
\hline
Number of Groups & 3,420\\
\hline
Size of the largest Group & 5,979\\
\hline
Number of Wishlist & 40,936\\
\hline
Size of the largest Wishlist & 5,892\\
\hline
Average Wishlist Size& 36.17\\
\hline \hline
\end{tabular}
\end{table}
}

From Table \ref{tab:main-anobii} it emerges that the number of users who joined at least one group is about 1.67 times higher than the number of users who applied at least one tag.
The percentage of users who exploited both tagging and group services were about 22.2\% of the whole user population.
About 40, 900 users produced a wishlist (consisting, on average, of 36.17 books.)

\section{Research Questions}\label{sec:researchquestions}
We formulate our set of research questions in the following:

\smallskip

\noindent
$\mathbf{Q_1}$. {\em What is the intensity of user activities?}
We are interested in measuring how many tags a user applied, how many groups she joined and how many books have been inserted in her wishlist.
Other studies from Social Web literature and social tagging systems \cite{de2010query,lux2007aspects,heymann2008can} inform us about the existence of few, prolific users who are responsible of most of the activities taking place in a Social Web platform.
We aim at checking whether such a behavior emerges also in aNobii and whether relevant differences emerge across different dimensions.

\smallskip

\noindent
$\mathbf{Q_2}$. {\em How do user perceive tagging activities?}
We are interested in analyzing the semantics associated with tags applied by aNobii user.
In particular, we want to check whether users prefer to apply generic tags having a broad meaning or, vice versa, if they privilege specialized tags with a narrow meaning.
We also look at checking whether tags are perceived as a knowledge management tool (and, therefore, they are mainly applied to classify books) or, vice versa, if they reflect personal user tastes.
To analyze tag semantics we will use Natural Language techniques and topic modeling.

\smallskip

\noindent
$\mathbf{Q_3}$. {\em Is there a correlation between the profiles of a user according to two different dimensions?}
We aim at observing users' behavior across different dimensions; the goal is to find out whether user behavior significantly differs across different dimensions.
In this way, we will be able to understand to what extent semantic behavior differs from social one: for example, we will investigate whether users who are heavily involved in tagging activities are also more likely to join groups or create long wishlists.

\smallskip

\noindent
$\mathbf{Q_4}$. {\em What is the information content of each user profile? For instance, are tag-based profiles more informative than group-based ones?}
In other words, do users prefer to focus on some tags exploited by few other aNobii users?
In case of affirmative answer we can conclude that the semantic behavior of a user can be more easily predicted on the basis of the behavior of other users of aNobii platform.
We can repeat such an analysis for social behavior by checking whether the groups a user will decide to join or the books she will insert in her wishlist can be predicted on the basis of the behavior of other users.
To perform such an analysis we will rely on Information Theory techniques.

\smallskip

\noindent
$\mathbf{Q_5}$ {\em Do extreme behaviors emerge in user behaviors?}
We are interested in performing a {\em cross analysis} of user behavior along different dimensions.
Our investigation targets at answering questions like these: are there users who are heavily involved in tagging activities and, simultaneously, joined very few groups?
Are there users who have compiled short wishlists but have applied a large number of tags?
Such an analysis is relevant to disclose how users perceive the services provided by the aNobii platform.
For example, if we discover that a fraction of users who joined a large number of groups but who did not applied tag at all (or in a limited fashion), then we could conclude that some users perceive the aNobii platform as a social networking Web site rather than a container for organizing their bookshelves and share it with other users.

\section{The intensity of user activities}\label{sec:userprofanalysis}

Our first series of experiments focuses on the structure of tag-based, group-based and wishlist-based user profiles.
We start discussing tag-based profiles and we compute the {\em empirical Cumulative Distribution Function} (CDF) associated with the size of tag-based profiles.
The CDF $F_X(x)$ specifies the proportion of elements in a distribution $X$ which are less than a threshold value $x$.
It ranges between $0$ and $1$ and it is a monotone non-decreasing function.

\begin{figure*}[t!]\centering
\subfigure[]{
 \label{fig:cdf-tag-based}
 \begin{minipage}[tb]{0.32\textwidth}
  \centering
  \includegraphics[width=\textwidth]{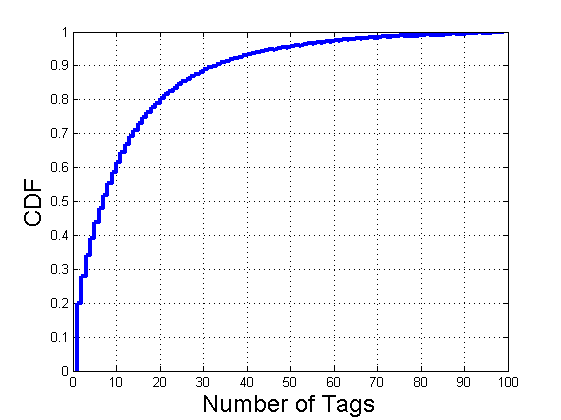}
   \end{minipage}}
\subfigure[]{
 \label{fig:cdf-group-based}
 \begin{minipage}[tb]{0.32\textwidth}
  \centering
  \includegraphics[width=\textwidth]{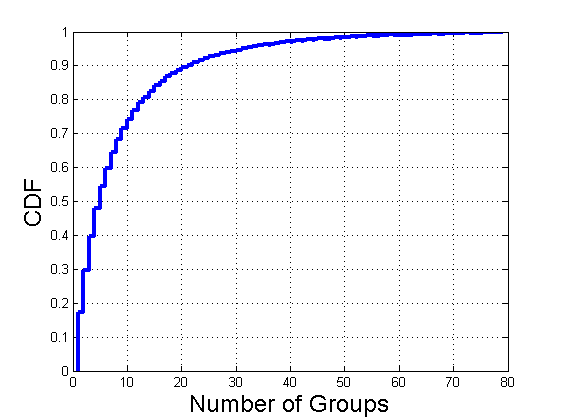}
   \end{minipage}}
\subfigure[]{
 \label{fig:cdf-wish-based}
 \begin{minipage}[tb]{0.32\textwidth}
  \centering
  \includegraphics[width=\textwidth]{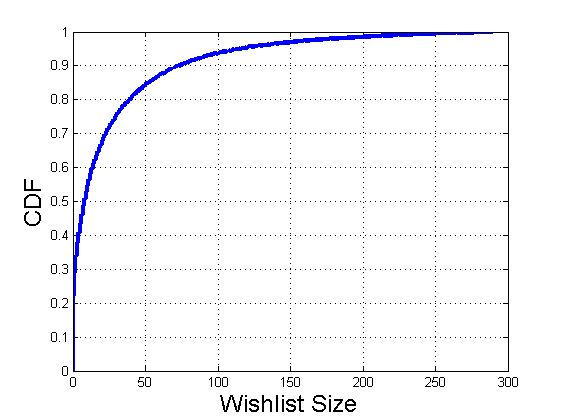}
   \end{minipage}}
 \vspace{-0.2cm}
\caption{Cumulative Distribution Function of tag-based, group-based and wishlist-based profiles size. The diagram on the left ({\bf a}) is about tag-based profiles, the diagram in the middle ({\bf b}) is about group-based profiles, and, finally, the diagram on the right ({\bf c}) is about wishlist-based profiles.}
\label{fig:cdf}
\end{figure*}

The CDF associated with tag-based profile size is reported in Figure \ref{fig:cdf-tag-based}.
For a given value $\overline{x}$, Figure \ref{fig:cdf-tag-based} allows to observe the probability that the size of the tag-based profile of a user is less than or equal to $\overline{x}$.
In an analogous fashion, we calculated the CDF associated with the size of group-based and wishlist-based profiles and the obtained results are plotted in Figures \ref{fig:cdf-group-based} and \ref{fig:cdf-wish-based}.

Some interesting observations can be drawn from these figures.
As for tag-based profiles, we know that the largest tag-based profile consisted of 2,622 tags, whereas the average size of tag-based profiles is 16.59 with a standard deviation equal to 46.08 and a median of 7.
Despite few exceptional values, we can conclude that users generally apply few tags and over 90\% of users applied less than 30 tags.
The adoption of tags in aNobii is therefore broadly distributed, and this is consistent with findings reported for traditional folksonomies \cite{lux2007aspects,Cattuto*07,halpin2007complex}.

An analogous trend emerges for group-based profiles (see Figure \ref{fig:cdf-group-based}.)
However, the number of groups users join is much smaller than the number of tags they use.
We found that the largest number of groups a user decided to join was equal to 554.
Most users joined around or less than 10 groups: the average size of group-based profiles was, in fact, 9.94 with a standard deviation equal to 17.49 and a median of 5.
Once again, the CDF describing the size of group-based profiles grows quite quickly telling that about 90\% of the user population joined less than 20 groups.
The dynamics of group formation and joining exhibited by aNobii closely resemble those reported in literature for other technologically-mediated social networks, \emph{e.g.,} \emph{LiveJournal} \cite{backstrom2006group}.

The most surprising result, however, comes from the analysis of wishlist-based profiles.
Although roughly 90\% of the users compiled wishlists containing at most 75 books, aNobii wishlists show a broad distribution as well (see Figure \ref{fig:cdf-wish-based}.)
This yields few users having very long wishlists (the maximum was 5,892) with an average size equal to 36.17, a standard deviation equal to 113.77 and a median of 9.

We note that the range of variability associated with the size of wishlist-based profiles is much larger than the size of tag-based and group-based profiles.

This suggests that there are some users who do not perceive wishlists as a useful tool and insert just few or no books in their wishlists.
Otherwise, a small fraction of users interprets wishlists as a relevant facility provided by the aNobii platform and provide a long list of books they are likely to read.

These results help us in answering $\mathbf{Q_1}$: in all the considered dimensions we found few prolific users who were heavily involved in tagging, joining groups and compiling wishlists, whereas the large amount of other users was mostly inactive or silent.
This implies that user profiles are, on average, quite sparse in all the considered dimensions.
However, the degree of variability in the size of user profiles is non-uniform across these dimensions and the highest level of variability emerges in the usage of wishlists.

%
%

\section{The role of tagging in aNobii}\label{sec:role-of-tagging}

Here we analyze the semantics of aNobii tags by means of {\em Latent Dirichlet Allocation} (or {\em LDA} in short) \cite{BlNgJo03}. It maps documents (modeled as a bag of words in a {\em high dimensional space}) onto points of a low-dimensional space called {\em topic space}. LDA can be seen as a dimensionality reduction technique such that each document may be viewed as a {\em mixture} of various topics. LDA resembles \emph{probabilistic latent semantic analysis} (pLSA), with the exception that in LDA the topic distribution is assumed to have a Dirichlet prior.

Each topic generates a collection of words and each word is associated with the probability of being observed in a document.
A word can occur (possibly with different probabilities) in different topics.
Topics can be viewed as abstract entities and, therefore, a topic may not have a clear interpretation and human experts could be asked to associate semantics with topics.
Suppose to consider a collection of documents related to a topic referring to travels.
This topic is likely to generate with high probability words like {\em Hotel}, {\em Railways} or {\em Airlines}.
The number of topics $N_{\mathrm{Topics}}$ that LDA has to discover is a parameters of the algorithm which must be provided; in real cases a reasonable choice is to fix $N_{\mathrm{Topics}}$ in the range 50-300.

LDA has been recently applied to folksonomies \cite{jin2011topic,SiSi09} being the collection of documents replaced by the set of items available in the folksonomy.
In case of aNobii, items coincide with books and each book is described as a bag of words whose elements are tags.
Each topic $z_j$ is associated with a {\em topic vector} $\vec{\mathbf{\theta}}_{z_j} \in \R^{|T|}$, a vector having as many components as the tags in $T$ and the $i$-th component $\vec{\mathbf{\theta}}_{z_j}[i]$ of $\vec{\mathbf{\theta}}_{z_j}$ specifies the probability that the topic $z_j$ generates the tag $t_i$.

\begin{table*}[htbp] \small
\label{tab:topic-distribution}
\centering
\caption{Five Topics extracted from aNobii folksonomy}
\subtable[Topic 1\label{tab:topic1}]{%
\begin{tabular}{l|l}
\multicolumn{1}{c|}{\textbf{Word}} &
\multicolumn{1}{c}{\textbf{Pr}} \\
\hline
Crime Novel & 0.549 \\
My Favorites & 0.144 \\
Hobby & 0.088 \\
Cats & 0.031 \\
Novel & 0.021 \\
Other & 0.167 \\
\end{tabular}
}
\subtable[Topic 2\label{tab:topic2}]{%
\begin{tabular}{l|l}
\multicolumn{1}{c|}{\textbf{Word}} &
\multicolumn{1}{c}{\textbf{Pr}} \\
\hline
USA & 0.313 \\
Adventure & 0.162 \\
Japan & 0.067 \\
Science & 0.064 \\
Dictionaries & 0.042 \\
World & 0.023 \\
Other & 0.329 \\
\end{tabular}
}
\subtable[Topic 3\label{tab:topic3}]{%
\begin{tabular}{l|l}
\multicolumn{1}{c|}{\textbf{Word}} &
\multicolumn{1}{c}{\textbf{Pr}} \\
\hline
Philosophy & 0.534 \\
Buddhism & 0.253 \\
Divulgation & 0.028 \\
East & 0.016 \\
Historical  & 0.013 \\
Other &  0.156 \\
\end{tabular}
}
\subtable[Topic 4\label{tab:topic4}]{%
\begin{tabular}{l|l}
\multicolumn{1}{c|}{\textbf{Word}} &
\multicolumn{1}{c}{\textbf{Pr}} \\
\hline
Cultural Studies &  0.472\\
Humanities & 0.133 \\
Beaux-Arts & 0.057 \\
Autobiography & 0.036\\
Biography & 0.035\\
Writing & 0.032 \\
Malaparte & 0.02 \\
Fun & 0.012 \\
Other & 0.203 \\
\end{tabular}
}
\subtable[Topic 5\label{tab:topic5}]{%
\begin{tabular}{l|l}
\multicolumn{1}{c|}{\textbf{Word}} &
\multicolumn{1}{c}{\textbf{Pr}} \\
\hline
French Literature &  0.591\\
Thriller & 0.108 \\
Legal & 0.106 \\
Books & 0.029\\
Satire & 0.013\\
Essay Writing & 0.012 \\
Society & 0.005 \\
Humor & 0.005 \\
Other & 0.148 \\
\end{tabular}
}
\end{table*}

We applied a fast implementation of LDA described in \cite{HoBaBl10} with $N_{\mathrm{Topics}} = 100$.
In Table \ref{tab:topic-distribution}, we show 5 example topics.
For each topic we report the words ({\textbf{Word}) it generates and the probability ({\textbf{Pr}) of generating that word.
Words are sorted in decreasing order of the probability of being generated.

The main facts emerging from the topical analysis follow:

\begin{itemize}


\item Tags are used to classify literary genres (\emph{e.g.,} think of {\em Crime Novel} in Table \ref{tab:topic1}, {\em Adventure} in Table \ref{tab:topic2} and {\em Thriller} in Table \ref{tab:topic5}.)

\item In some cases, tags have a narrow meaning and are used to classify books belonging to some specific research areas (\emph{e.g.,} {\em Cultural Studies} and {\em Humanities} in Table \ref{tab:topic4} or {\em Philosophy} and {\em Buddhism} in Table \ref{tab:topic3}.)
    Generic tags, however, may appear in conjunction with other tags that contribute to better specify them (\emph{e.g.,} {\em French Literature} is paired with more specific tags like {\em Satire}, {\em Legal}, {\em Society} or {\em Humor}.)

\item Tags are used {\em not only} with the goal of describing book features but also to state the personal viewpoint of a user.
In particular, think of the tag {\em My Favorite} appearing in Table \ref{tab:topic1} or the tag {\em fun} in Topic 4, Table \ref{tab:topic4}.

\item In some cases, tags are used not with the goal of describing book features neither to express user tastes: this is, for instance, the case of the {\em Malaparte} in Topic 3, being C. Malaparte, a famous Italian short-story writer and novelist.
   In such a case, in fact, a tag links a book with its authors; in other cases (not reported in Table \ref{tab:topic-distribution}), tags are used to specify the publication year of a book.
\end{itemize}

In light of these results, we are now able to answer $\mathbf{Q}_2$: the usage of tags in aNobii is very heterogeneous, because tags can have a narrow meaning and focus on a specific literary genre (sometimes tags are used to associate a book with its author); by contrast, tags can have a broad semantic (for instance, some users are likely to apply generic tags like {\em French} or {\em Russian Literature} which can refer to a very large collection of books.)
Tags may also be used to reflect user preferences rather than being used to categorize books.

\section{Correlation between user profiles across different dimensions} \label{sec:comparingdimensions}

We now discuss the comparison of the profiles of the same user across different dimensions.
We adopt the following procedure: {\em (i)} for each user $u_i \in U$ we consider her profiles $\vec{p}_T(i)$, $\vec{p}_G(i)$ and $\vec{p}_W(i)$; {\em (ii)} we take these profiles is a pairwise fashion and compute their Spearman's $\rho$ coefficient.
This yielded three different {\em configurations}.

The Spearman's $\rho$ coefficient is a non-parametric measure of dependence of two variables and it is used to assess whether one of the two variables can be described as a monotonic function of the other one.
In detail, given two vectors $\vec{x}$=$\{x_1,\ldots, x_n\}$ and $\vec{y}$=$\{y_1,\ldots, y_n\}$, the computation of $\rho$ requires to convert each component $x_i$ (resp., $y_i$) onto an ordinal value $rx_i$ (resp., $ry_i$), called \textit{rank}, such that the largest component of $\vec{x}$ (resp., $\vec{y}$) has rank 1 and the smallest one of $\vec{x}$ (resp., $\vec{y}$) has rank $n$.
This yields two variables $\vec{X} = \{rx_1,\ldots, rx_n\}$ and $\vec{Y} = \{ry_1,\ldots, ry_n\}.$
The $\rho$ coefficient is computed as the Pearson coefficient of $\vec{X}$ and $\vec{Y}$.
We sorted users according to their values of $\rho$ and the results are graphically reported in Figure \ref{fig:spearman}.

\begin{figure}[!t]
\includegraphics[scale=0.6]{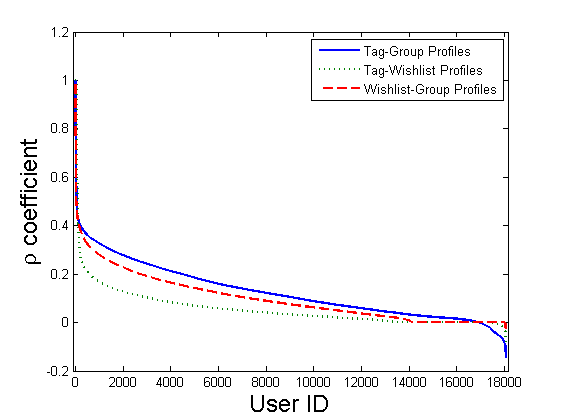}
\caption{Distribution of Spearman' $\rho$ coefficient}
\label{fig:spearman}
\end{figure}

From Figure \ref{fig:spearman} we observe that for a large part of user population (about 98.9\%) the value of $\rho$ is less than 0.4: this denotes a low level of correlation between tag-based, group-based and wishlist-based profiles.
The highest values of $\rho$ are achieved when we compare tag-based and group-based profiles, the lowest ones instead occur when we compare tag-based and wishlist based profiles.
This implies that the different profiles we may associate with a user provide information which complement each other and, taken as a whole, they allow for a better description of user behaviors.

\begin{figure*}[t!]
\subfigure[]{
 \label{fig:perm-tg}
 \begin{minipage}[tb]{0.33\textwidth}
  \centering
  \includegraphics[width=\textwidth]{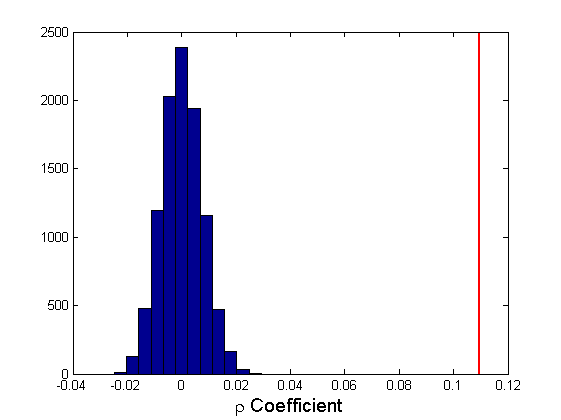}
   \end{minipage}}
\subfigure[]{
 \label{fig:perm-tl}
 \begin{minipage}[tb]{0.33\textwidth}
  \centering
  \includegraphics[width=\textwidth]{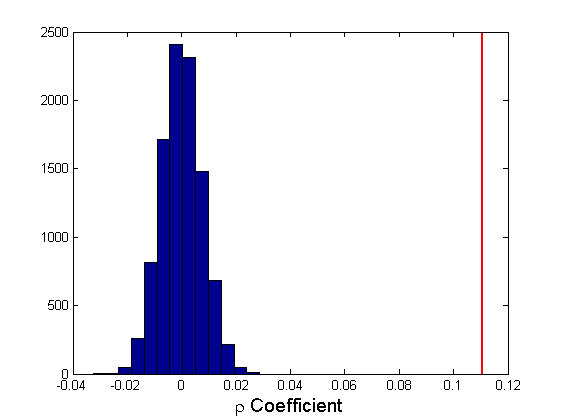}
   \end{minipage}}
\subfigure[]{
 \label{fig:perm-lg}
 \begin{minipage}[tb]{0.33\textwidth}
  \centering
  \includegraphics[width=\textwidth]{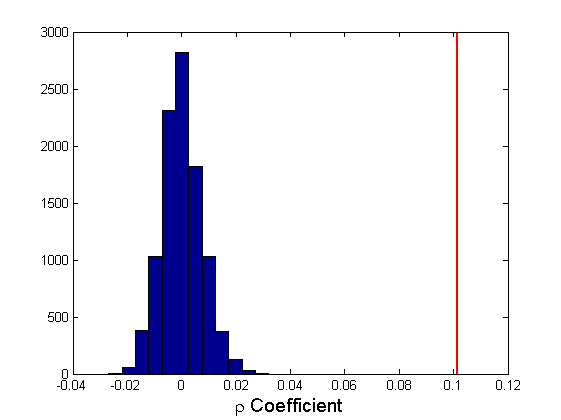}
   \end{minipage}}
 \vspace{-0.2cm}
\caption{Results of a permutation test for checking if the Spearman's $\rho$ coefficient is significant ($H_0: \rho = 0$.) The diagram on the left ({\bf a}) is about tag-based and wishlist-based profiles, the diagram in the middle ({\bf b}) is about tag-based and wishlist-based profiles, and, finally, the diagram on the right ({\bf c}) is about group-based and wishlist-based profiles. In all the diagrams the red line specifies the observed value of $\rho$.}
\label{fig:permutation}
\end{figure*}

To substantiate the statistical validity of the results, we check whether the $\rho$ coefficients are significantly different from 0.
To this purpose, we use a {\em permutation test} \cite{good2005permutation}, which is a re-sampling procedure allowing us to estimate the probability of obtaining a particular value of $\rho$ by chance.
Suppose to consider a pair of user profiles referring to a particular user $u_i$, say her tag-based and group-based profiles; the {\em null hypothesis} assumes that the Spearman's $\rho$ of $\vec{p}_T(i)$ and $\vec{p}_G(i)$ is 0, \emph{i.e.,} $H_0: \rho(\vec{p}_T(i),\vec{p}_G(i)) = 0$ whereas the alternative hypothesis is $H_a: \rho(\vec{p}_T(i),\vec{p}_G(i)) \neq 0$.
Let $\overline{\rho}$ be the observed value of the Spearman's correlation coefficient.

The permutation test requires to fix the tag-based profile $\vec{p}_T(i)$ of $u_i$ and to select, uniformly at random, a user $u_j$; after that, the Spearman's $\rho$ coefficient between $\vec{p}_T(i)$ and $\vec{p}_G(j)$ is computed.
The procedure described above is called {\em re-shuffling}: this procedure should be iterated a sufficiently high number of times (say $N_{\mathrm{reps}}$) to guarantee statistically robust results.
At each iteration we obtain a specific value of the Spearman's $\rho$ coefficient and, after many re-shuffling iterations we can plot a histogram reporting the distribution of $\rho$ values along with the observed value $\overline{\rho}$.
If $\overline{\rho}$ is far apart the tail of the histogram, we can reject the null hypothesis.
We can finally compute the p-value as the number of observed correlation values exceeding $\overline{\rho}$: if it is lower than a significance value $\alpha$ we have strong evidence that $\overline{\rho}$ differs from 0.

The re-shuffling procedure is applied to the three possible combinations of tag-, group- and wishlist-based profiles.
In our experiments we fixed $N_{\mathrm{reps}} = 10,000$.
Since the number of tests to carry out is very large, we focus only on few specific (but relevant) cases in which $\rho$ is around 0.1.
These users are the majority of the population in our dataset.
We report the result of our permutation test for users showing a value of $\rho$ equal to 0.1097 (tag-based and group-based profiles), 0.1016 (tag-based and wishlist-based profiles) and 0.1107 (group-based and wishlist-based profiles.)
The obtained results are reported in Figures \ref{fig:perm-tg}, \ref{fig:perm-tl} and \ref{fig:perm-lg}. In all cases, the p-value was less than $10^{-5}$: this tells us that $\rho$ coefficients are significantly different from 0.

From Figure \ref{fig:spearman} we also note that the lowest value of $\rho$ is about -0.17 and it occurs in case of tag-based and group-based profiles.
The fraction of user profiles showing a negative correlation is around 5.25\% in all the three configurations under investigation.
The fact that there exists a small negative correlation for a fraction of tag-based profiles paired with group- or wishlist-based ones, suggests that the broad distribution of tag usage may represent a confounding factor, thus making tag usage a bad predictor of overall user behavior.

We are now able to answer $\mathbf{Q}_3$: the correlation between user profiles is in general quite small, independently of the dimensions we consider.
This result suggests that we should integrate information available in each profile to get a more detailed picture of users' behavior and needs.

\section{Entropy of User Profiles}\label{sub:analysis-entropy}
In this section we assess the value of information embodied in aNobii user profiles.
In this way, we can address a fundamental question: to what extent can future user activities be predicted on the basis of the behavior of other aNobii users?
We address this question by means of the application of Shannon's {\em entropy measure}~\cite{shannon48} to tag-based profiles.
We first define the probability that users $u_i$ and $u_j$ applied the same tags as

\begin{eqnarray}\label{eq:entropy-tag-prop}
\textrm{Pr}_{T}(i,j) = \frac{\vec{p}_T[i,j]}{\sum_{l \in U} \vec{p}_{T}[i,l]}
\end{eqnarray}

\noindent
Then, the entropy $\mathcal{E}(\vec{p}_T(i))$ of $\vec{p}_T(i)$ is defined as follows

\begin{equation}\label{eq:entropy}
\mathcal{E}(\vec{p}_T(i))= -\sum_{u_j \in U} \textrm{Pr}_T(i,j) \cdot \log_2\left(\textrm{Pr}_T(i,j)\right).
\end{equation}

The term $-\log_2(\textrm{Pr}_T(i,j)$ is called \emph{self-information} and, in case $\textrm{Pr}_T(i,j) = 0$ we conventionally set $\textrm{Pr}_T(i,j) \cdot \log_2\left(\textrm{Pr}_T(i,j)\right) = 0$.

The entropy of $\vec{p}_T(i)$ specifies to what extent the tagging behavior of user $u_i$ is related to that of other aNobii users.
To clarify this concept, suppose that $u_i$ applied {\em exactly one tag} (say $\overline{t}$) and that $\overline{t}$ has been used also by user $u_j$.
In this case, we have $\vec{p}_T[i,j] = 1$ and $\vec{p}_T[i,l] = 0$ for each $l \neq j$.
This implies that $\log_2\left(\textrm{Pr}_T(i,j)\right) = 0$ and $\mathcal{E}(\vec{p}_T(i)) = 0$.
If, vice versa, $u_i$ applied multiple tags, say $t_1, \ldots, t_m$ that have been also applied by other users, then some components of $\vec{p}_T(i)$ will be greater than 0 and the value of $\mathcal{E}(\vec{p}_T(i))$ will be strictly greater than 0.
In such a case, the tag-based profile of $u_i$ exhibits a high degree of variability and it will be more difficult to predict what tags user $u_i$ will adopt in the future knowing the tags other users applied.
The entropy of one's is also influenced by its {\em size} and in the extreme case of an empty profile the associated entropy would be equal to 0.

The same line of reasoning holds for group-based and wishlist-based profiles.
Thus, we can generalize Equation \ref{eq:entropy} to define entropy for group-based and wishlist-based profiles as well.
The lower the entropy of a user profile, the less ``valuable'' the information in the profile, since future actions can be predicted on the basis of available information.

In Figures \ref{fig:entropy-scatter}(a)-(c) we show three scatter plots.
In each diagram we fixed two dimensions: a dot in each plot is uniquely associated with a user and its coordinates identify the entropy (expressed in bits) associated with the profile of that user in each dimension. For sake of interpretation, we also sorted users on the basis of their profiles entropy and we graphically reported the obtained results in Figure \ref{fig:entropy-profiles}.

\begin{figure*}[t!]
\subfigure[Tag-Based and Group-Based Profiles]{
 \label{fig:entropy-fd}
 \begin{minipage}[tb]{0.33\textwidth}
  \centering
  \includegraphics[width=\textwidth]{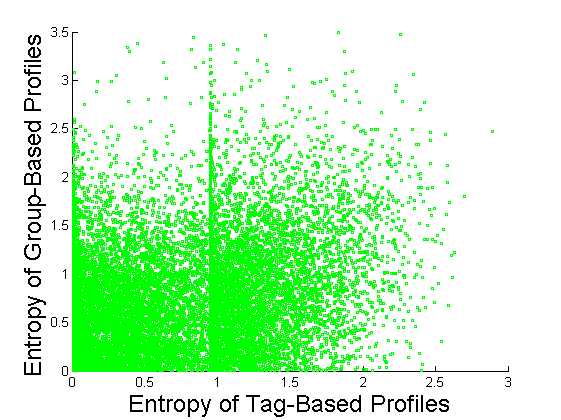}
   \end{minipage}}
\subfigure[Tag-Based and Wishlist-Based profiles]{
 \label{fig:entropy-fs}
 \begin{minipage}[tb]{0.33\textwidth}
  \centering
  \includegraphics[width=\textwidth]{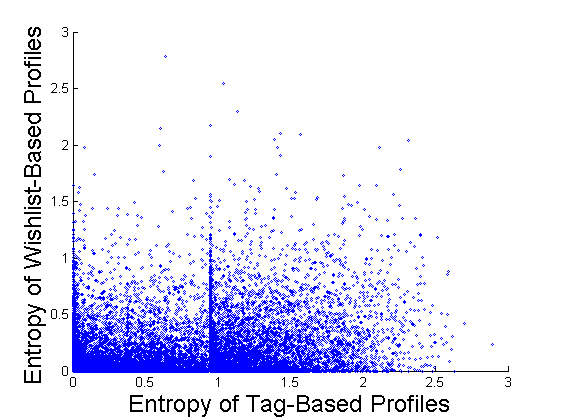}
   \end{minipage}}
\subfigure[Group-Based and Wishlist-Based Profiles]{
 \label{fig:entropy-sd}
 \begin{minipage}[tb]{0.33\textwidth}
  \centering
  \includegraphics[width=\textwidth]{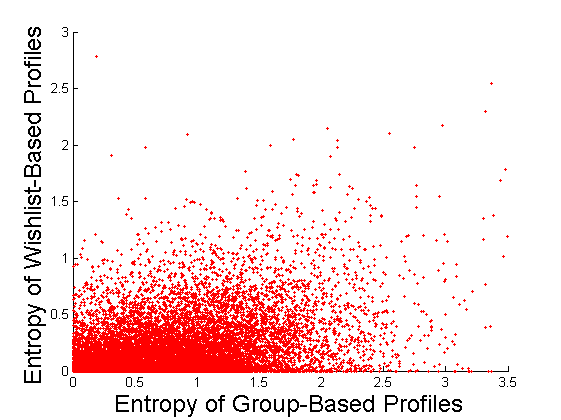}
   \end{minipage}}
 \vspace{-0.2cm}
\caption{Entropy of user profiles (in bits.) Each dot represents a user and shows the entropy of her profile in two out of the the three considered dimensions.}
\label{fig:entropy-scatter}
\end{figure*}

\begin{figure}[ht]
\centering
\includegraphics[scale=0.6]{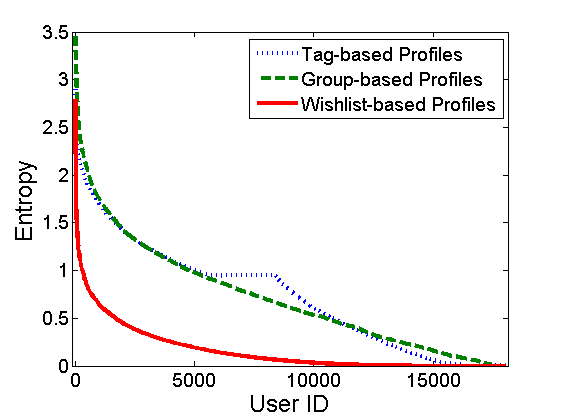}
\caption{Entropies of tag-based, group-based and wishlist-based user profiles}
\label{fig:entropy-profiles}
\end{figure}

Figures \ref{fig:entropy-scatter}(a)-(c) and Figure \ref{fig:entropy-profiles} provide us relevant hints to answer $\mathbf{Q}_3$.
From Figure \ref{fig:entropy-scatter}(a) we observe the presence of few users showing high level of entropy in their group-based profiles (around 3.5 bits) and low values of entropy in their tag-based profiles (around 0.5 bits.)
Conversely, these users are balanced by other users who feature high entropy values in their tag-based profiles and low values of entropy in their group-based profiles.
Users exhibiting a high level of discrepancy between the entropy associated with their profiles are less numerous if we consider, in a pairwise fashion, the $T$ and $W$ dimensions as well as the $G$ and $W$ dimensions.
However, independently of the pair of dimensions we are take into account, there is a relevant fraction of user population who shows relatively large values of entropy in the profile associated with the first dimension and relatively low values of entropy associated with the profile in the second dimension.
Therefore, the level of variability associated with a user profile is non-uniform across all dimensions.

An important observation from Figure \ref{fig:entropy-profiles} is that the entropy of tag-based and group-based user profiles turns out to be nearly comparable and, therefore, the information content of tag-based profiles is at least comparable to that of group-based ones.
Note that this is true at the aggregated level, while at the single user level it might be still possible that either the tag-based or the group-based profile convey more information than the others.
Such result suggests that, overall, group affiliation is as much relevant to disclose user interests as tagging behavior.
One possible explanation is the fact that aNobii users form groups on the basis of specific, shared reading interests and, therefore, the affiliation to a group is relevant to disclose user reading preferences.

The usage of tags is also interesting: on the one hand, aNobii provides category-based tags like {\em Science Fiction} or {\em Fantasy} allowing users to classify the content they produce in predefined classes.
Other social systems like {\em Delicious} or {\em Flickr} support collaborative tagging but, differently from aNobii, those platforms only employ {\em free-tagging}, \emph{i.e.,} users create and use the tags they prefer.
In aNobii, in addition to tags describing the main features of an object, we can find tags describing user personal moods (like {\em happy}) or personal events (like {\em graduation}) or even dates (\emph{e.g.,} a tag like $2008$ used to associate an object with the publication year.)
The usage of category-based tags lowers the degree of entropy of user profiles with respect to a free-tagging configuration by reducing the variability in the tag usage.
However, also aNobii users are free to create their own tags, which do not necessarily aim at describing a resource.
Those tags are shared by very few users and this as a result raises the level of entropy.
The lowest level of entropy is achieved by wishlist-based profiles, which are then less informative than tag- and group-based ones.

We now study the {\em mutual dependence} of information embodied in user profiles.
To this aim, we use a further measure from Information Theory known as {\em mutual information - MI} \cite{cover1991elements}.
Suppose that $X$ (defined over a domain $\mathcal{X}$) and $Y$ (defined over a domain $\mathcal{Y}$) are two random variables; the mutual information $I(X;Y)$ between $X$ and $Y$ specifies how much the knowledge of $X$ tells us about $Y$.
Let $p(x,y)$ be the joint probability distribution function of $X$ and $Y$; let $p(x)$ (resp., $p(y)$) be the marginal probability distribution functions of $X$ (resp., $Y$.)
Mutual information is defined as follows

\begin{equation}
\label{eqn:mutual}
I(X; Y) = \sum_{x \in \mathcal{X}}\sum_{y \in \mathcal{Y}} p(x,y) \log\left(\frac{p(x,y)}{p(x)p(y)}\right).
\end{equation}

The mutual information $I(X;Y)$ is non negative for any pair of random variables $X$ and $Y$ and it is also {\em symmetric}, \emph{i.e.,} $I(X; Y) = I(Y; X)$.
The higher $I(X; Y)$, the less uncertainty there is in determining $X$ (resp., $Y$) given that we know $Y$ (resp., $X$.)
If we adopt logarithms in base 2, then mutual information is measured in bits.

\begin{figure*}[t!]
\begin{minipage}[tb]{\textwidth}
\end{minipage}
\subfigure[Tag-Based and Group-Based Profiles]{
 \label{fig:Mutual-T-G}
 \begin{minipage}[tb]{0.33\textwidth}
  \centering
  \includegraphics[width=\textwidth]{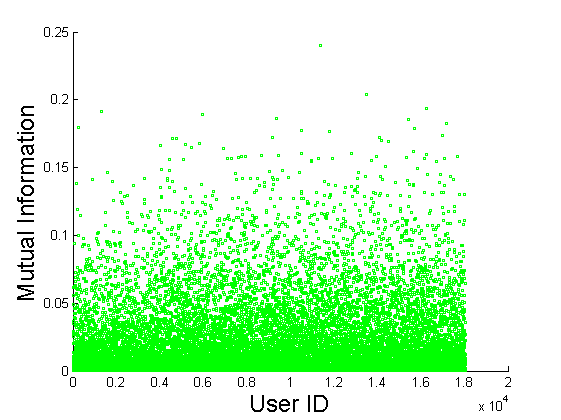}
   \end{minipage}}
\subfigure[Tag-Based and Wishlist-Based profiles]{
 \label{fig:Mutual-T-W}
 \begin{minipage}[tb]{0.33\textwidth}
  \centering
  \includegraphics[width=\textwidth]{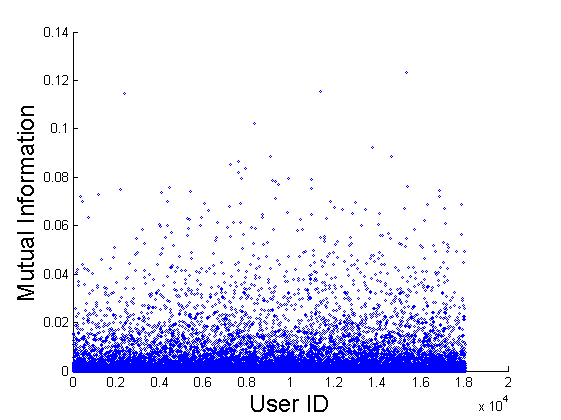}
   \end{minipage}}
\subfigure[Group-Based and Wishlist-Based Profiles]{
 \label{fig:Mutual-G-W}
 \begin{minipage}[tb]{0.33\textwidth}
  \centering
  \includegraphics[width=\textwidth]{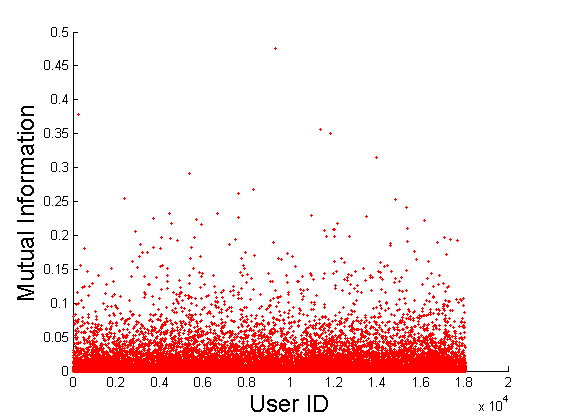}
   \end{minipage}}
 \vspace{-0.2cm}
\caption{Mutual information of user profiles (in bits.)}
\label{fig:mutual-information}
\end{figure*}

Equation \ref{eqn:mutual} easily generalizes to the case of arbitrary distributions: for a fixed user $u_i \in U$ we can consider tag-based, group-based and wishlist-based profiles in a pairwise fashion and, for each of these pairs, we can compute their mutual information.
As a result, each user $u_i$ is associated with three mutual information values.
In Figure \ref{fig:mutual-information}(a) we show each user ID in our dataset ($x$-axis) and the mutual information associated with tag-based and group-based profiles ($y$-axis.)
We also show the values of mutual information associated with tag-based and wishlist-based profiles (Figure \ref{fig:mutual-information}(b)) and group-based and wishlist-based profiles (Figure \ref{fig:mutual-information}(c).)

In general, the values of mutual information are quite low in all of the three configurations plotted in Figure \ref{fig:mutual-information}(a)-(c).
The worst case occurs for the mutual information associated with the tag-based and wishlist-based profiles and, for the vast majority of available users their tag-based and wishlist-based profiles seem independent (therefore, the knowledge of the content of the tag-based profiles is not effective in determining the content of her wishlist-based profile.)
There is, however, a small fraction of users such that the mutual information between their group-based and wishlist-based user profiles is greater than 0.2 (with a peak around 0.5.)

As a result of this analysis, we are now able to answer to $\mathbf{Q_4}$: tag-based and group-based profiles are in general more informative than wishlist-based ones.
Nevertheless, note that these profiles, taken independently, convey little information.
This is due to the fact that there exists a relevant fraction of users whose entropy in one profile is sensibly higher than in the others, therefore the variability associated with each user is not uniformly distributed across all profiles.
Concluding, different profiles convey different information: taken in pairs, the profiles share little mutual information, which means that the knowledge of only one dimension (\emph{i.e.,} profile) of a user is not particularly helpful to predict her behavior in the others.

\section{Multidimensional Cross Analysis}\label{sec:joint-tagging-group}
We conclude our analysis investigating contrasting behavior that emerges across different aNobii interaction channels.
We are interested in users who apply very few tags while, at the same time, join a large number of groups; another question of interest is whether there is a non-negligible fraction of aNobii users who produce rich wishlists but refrain from tagging their books.
Such analysis shall disclose how users perceive the features provided by the aNobii platform and what are the services they feel most comfortable.
To this aim, we consider the joint behavior of aNobii users across multiple dimensions and we adopt a suitable probability distribution function modeling user behavior across selected dimensions. To compute these joint probabilities, we employ a standard Bivariate Gaussian Kernel \cite{scott2009multivariate} and the corresponding results are reported as contour plots in Figures \ref{fig:biv1}, \ref{fig:biv2} and \ref{fig:biv3}.

\begin{figure}[t]
\centering
\includegraphics[scale=0.27]{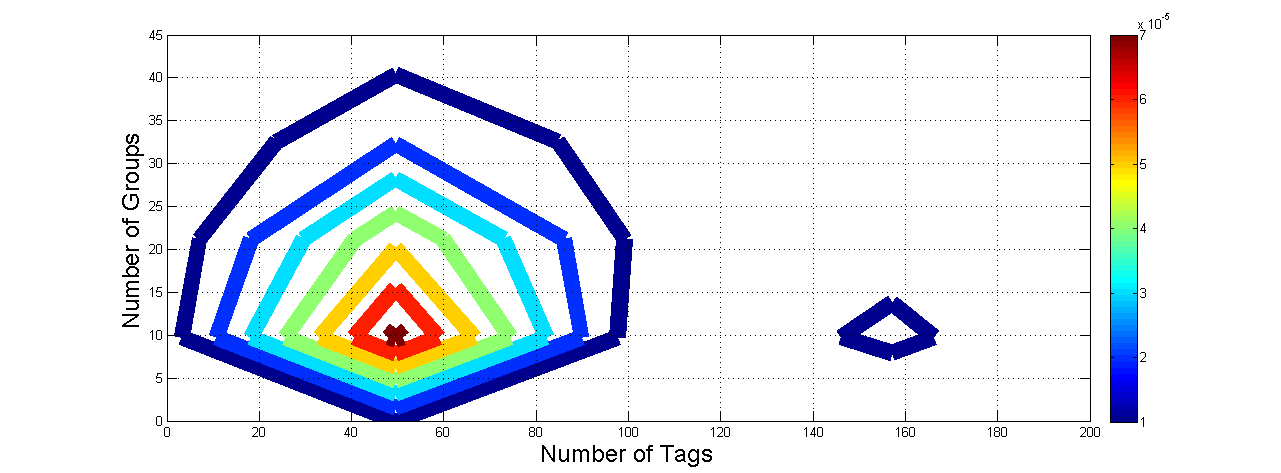}
\caption{Joint PDF associated with $T$ and $G$ dimensions}
\label{fig:biv1}
\end{figure}

\begin{figure}[t]
\centering
\includegraphics[scale=0.27]{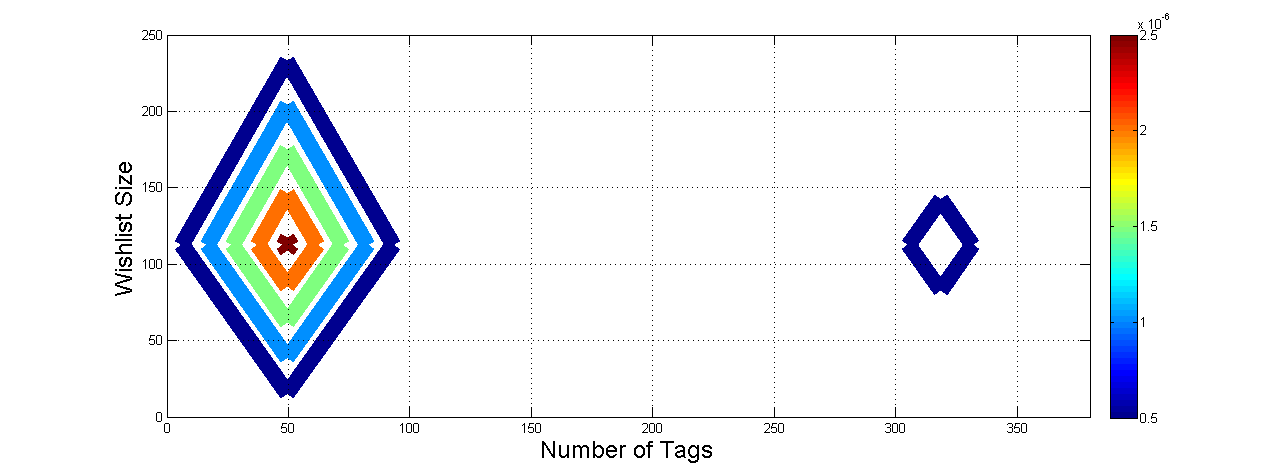}
\caption{Joint PDF associated with $T$ and $W$ dimensions}
\label{fig:biv2}
\end{figure}

\begin{figure}[t]
\centering
\includegraphics[scale=0.27]{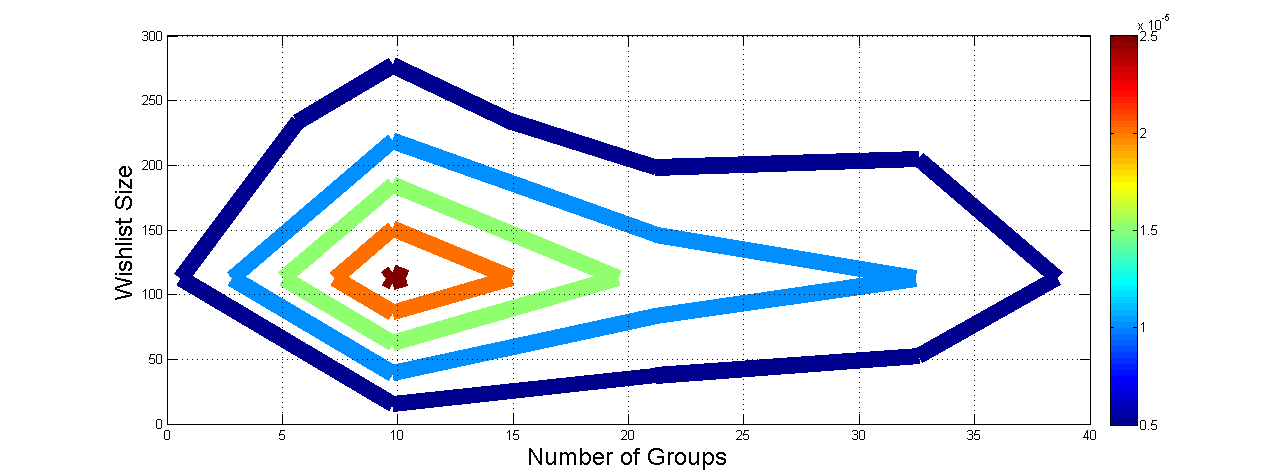}
\caption{Joint PDF associated with $G$ and $W$ dimensions}
\label{fig:biv3}
\end{figure}

As for $T$ and $G$ dimensions (see Figure \ref{fig:biv1}) we can notice two peaks: the stronger one is located near the point $(45, 10)$ whereas the weaker one is at the location $(150, 10)$.
The latter peak proves the presence of a little (but not negligible) fraction of user population who applied a relatively large number of tags but joined a relatively small number of groups.

As for $T$ and $W$ dimensions (see Figure \ref{fig:biv2}) we observe, once again, the presence of two peaks, the former roughly located at $(50, 110)$ and the latter at $(320, 120)$.
The second peak is quite interesting because it informs us about a non negligible fraction of users who applied a number of tags which ranges between 300 and 330, a value quite larger than the average number of tags aNobii users generally adopt.
However, the size of the wishlists associated with these users is rather small in comparison with the size of the largest wishlist available in our dataset.
There are some users who perceive tags as a powerful tool to organize their book collections as well as to expose them to the other aNobii users but who show less interest and motivations in compiling their wishlists.
Observe that the two regions in the contour plot in Figure \ref{fig:biv2} are symmetric and if we spin around $T$ and $W$ axes the contour plot looks the same.

Finally, as for $G$ and $W$ dimensions (see Figure \ref{fig:biv3}), there is a peak occurring at the point $(10, 120)$ but it is interesting to observe that the range of variability in the number of groups a user joined to is quite high (it ranges from 0 to about 40.)

We answer question $\mathbf{Q}_5$ by observing that extreme behavior actually exists and this discrepancy emerges especially in the usage of tags, although not very commonly.
The majority of aNobii users exhibit a balanced adoption of the features of the platform: either they are mostly inactive in all dimensions, or they are engaged in multiple activities to some extent.

\section{Conclusions}\label{sec:conclusions}
In this work we analyzed the aNobii social network of book readers, which allows users to post their readings, give ratings, review books and discuss them with friends and fellows.
This environment is interesting due to its heterogeneous structure, part social network, part interest network and part folksonomy.
We carried out an extensive analysis describing aNobii users profiles according to three different dimensions, seeking to understand what type of patterns governs the usage of tags, the joining of groups and the compilation of reading wishlists.

From our analysis it emerges that these three activities are described by broad distributions: many users (roughly 90\%) exhibit moderately low levels of participation to the platform activities, but the remainder of users is increasingly active.
We investigated whether any form of correlation between these three activities exists, discovering that the correlation is usually quite small, suggesting the need to incorporate multiple dimensions to effectively describe users profiles.

We also noticed that the information encoded in the tag-based and group-based profiles is more valuable than that of wishlist-based ones.
Each dimension is described by its own entropy distribution that, if taken independently, does not convey much information; in fact, the mutual information between pairs of dimensions is low.
This is due to the fact that a relevant fraction of users exhibits unbalanced entropy across different dimensions, with one's profile entropy sensibly higher than other profiles.
This suggests that the variability associated with each user is not uniformly distributed across all dimensions, and it leads to hypothesize that these dimensions complement each other, even if to different extents.
We concluded our analysis noticing that extreme behavior (\emph{e.g.,} users very active in one dimension and silent in the others) emerges but not very commonly, with the majority of users showing a balance in the adoption of all three observed behavior.

%
%
%

As for future work, we plan to design, implement and experimentally validate a recommender algorithm capable of suggesting readers new books on the basis of both social and semantic signals, exploiting the above-mentioned user dimensions explored throughout the paper.

\section*{Acknowledgments}
The authors are grateful to Luca M. Aiello for providing access to the aNobii dataset object of this study.
\bibliographystyle{IEEETran}
\bibliography{IEEEabrv,../../anobii-bibliography}

\begin{thebibliography}{10}
\providecommand{\url}[1]{#1}
\csname url@samestyle\endcsname
\providecommand{\newblock}{\relax}
\providecommand{\bibinfo}[2]{#2}
\providecommand{\BIBentrySTDinterwordspacing}{\spaceskip=0pt\relax}
\providecommand{\BIBentryALTinterwordstretchfactor}{4}
\providecommand{\BIBentryALTinterwordspacing}{\spaceskip=\fontdimen2\font plus
\BIBentryALTinterwordstretchfactor\fontdimen3\font minus
  \fontdimen4\font\relax}
\providecommand{\BIBforeignlanguage}[2]{{%
\expandafter\ifx\csname l@#1\endcsname\relax
\typeout{** WARNING: IEEEtran.bst: No hyphenation pattern has been}%
\typeout{** loaded for the language `#1'. Using the pattern for}%
\typeout{** the default language instead.}%
\else
\language=\csname l@#1\endcsname
\fi
#2}}
\providecommand{\BIBdecl}{\relax}
\BIBdecl

\bibitem{santos2009individual}
E.~Santos-Neto, D.~Condon, N.~Andrade, A.~Iamnitchi, and M.~Ripeanu,
  ``Individual and social behavior in tagging systems,'' in \emph{Proc.
  International Conference on Hypertext and Hypermedia}, 2009.

\bibitem{Aiello*10}
L.~M. Aiello, A.~Barrat, C.~Cattuto, G.~Ruffo, and R.~Schifanella, ``{Link
  Creation and Profile Alignment in the aNobii Social Network},'' in
  \emph{Proc. International Conference on Social Computing}, 2010, pp.
  249--256.

\bibitem{aiello2012people}
L.~M. Aiello, M.~Deplano, R.~Schifanella, and G.~Ruffo, ``People are strange
  when you’re a stranger: Impact and influence of bots on social networks,''
  in \emph{Proc. 6th International Conference on Weblogs and Social Media},
  2012, pp. 10--17.

\bibitem{Aiello*13}
L.~M. Aiello, A.~Barrat, C.~Cattuto, R.~Schifanella, and G.~Ruffo, ``{Link
  creation and information spreading over social and communication ties in an
  interest-based online social network},'' \emph{EPJ Data Science}, 2013.

\bibitem{demeo2013analyzing}
P.~D. Meo, E.~Ferrara, F.~Abel, L.~Aroyo, and G.~Houben, ``Analyzing user
  behavior across social sharing environments,'' \emph{ACM Transactions on
  Intelligent Systems and Technology}, vol.~5, no.~1, 2013.

\bibitem{zhou2010solving}
T.~Zhou, Z.~Kuscsik, J.~Liu, M.~Medo, J.~Wakeling, and Y.~Zhang, ``Solving the
  apparent diversity-accuracy dilemma of recommender systems,''
  \emph{Proceedings of the National Academy of Sciences}, vol. 107, no.~10, pp.
  4511--4515, 2010.

\bibitem{DeMeo*11}
P.~{De Meo}, A.~Nocera, G.~Terracina, and D.~Ursino, ``{Recommendation of
  similar users, resources and social networks in a Social Internetworking
  Scenario},'' \emph{Information Sciences}, vol. 181:7, pp. 1285--1305, 2011.

\bibitem{ma2011recommender}
H.~Ma, D.~Zhou, C.~Liu, M.~Lyu, and I.~King, ``Recommender systems with social
  regularization,'' in \emph{Proc. International Conference on Web Search and
  Data Mining}.\hskip 1em plus 0.5em minus 0.4em\relax ACM, 2011, pp. 287--296.

\bibitem{kazienko2011multidimensional}
P.~Kazienko, K.~Musial, and T.~Kajdanowicz, ``Multidimensional social network
  in the social recommender system,'' \emph{{IEEE} Trans. Syst., Man, Cybern.
  A, Syst., Humans}, vol.~41, no.~4, pp. 746--759, 2011.

\bibitem{centola2010spread}
D.~Centola, ``The spread of behavior in an online social network experiment,''
  \emph{Science}, vol. 329, no. 5996, pp. 1194--1197, 2010.

\bibitem{lerman2010information}
K.~Lerman and R.~Ghosh, ``Information contagion: An empirical study of the
  spread of news on digg and twitter social networks.'' in \emph{Proc.
  International Conf. on Weblogs and Social Media}, 2010, pp. 90--97.

\bibitem{weng2012competition}
L.~Weng, A.~Flammini, A.~Vespignani, and F.~Menczer, ``Competition among memes
  in a world with limited attention,'' \emph{Sci. Rep.}, 2012.

\bibitem{conover2013geospatial}
M.~Conover, C.~Davis, E.~Ferrara, K.~McKelvey, F.~Menczer, and A.~Flammini,
  ``{The geospatial characteristics of a social movement communication
  network},'' \emph{PloS one}, vol.~8, no.~3, p. e55957, 2013.

\bibitem{ferrara2011topological}
E.~Ferrara and G.~Fiumara, ``Topological features of online social networks,''
  \emph{Communications in Applied and Industrial Mathematics}, vol.~2, no.~2,
  pp. 1--20, 2011.

\bibitem{ferrara2012large}
E.~Ferrara, ``{A large-scale community structure analysis in Facebook},''
  \emph{EPJ Data Science}, vol.~1, no.~9, pp. 1--30, 2012.

\bibitem{grabowicz2012social}
P.~Grabowicz, J.~Ramasco, E.~Moro, J.~Pujol, and V.~Eguiluz, ``Social features
  of online networks: The strength of intermediary ties in online social
  media,'' \emph{PLoS ONE}, vol.~7, no.~1, p. e29358, 2012.

\bibitem{conover2013digital}
M.~D. Conover, E.~Ferrara, F.~Menczer, and A.~Flammini, ``The digital evolution
  of occupy wall street,'' \emph{PloS one}, vol.~8, no.~5, 2013.

\bibitem{mcpherson2001birds}
M.~McPherson, L.~Smith-Lovin, and J.~Cook, ``Birds of a feather: Homophily in
  social networks,'' \emph{Annu. Rev. Sociol.}, pp. 415--444, 2001.

\bibitem{aiello2012friendship}
L.~Aiello, A.~Barrat, R.~Schifanella, C.~Cattuto, B.~Markines, and F.~Menczer,
  ``Friendship prediction and homophily in social media,'' \emph{ACM
  Transactions on the Web}, vol.~6, no.~2, p.~9, 2012.

\bibitem{bollen2011happiness}
J.~Bollen, B.~Gon{\c{c}}alves, G.~Ruan, and H.~Mao, ``Happiness is assortative
  in online social networks,'' \emph{Artificial Life}, vol. 17:3, pp. 237--251,
  2011.

\bibitem{golder2011diurnal}
S.~Golder and M.~Macy, ``Diurnal and seasonal mood vary with work, sleep, and
  daylength across diverse cultures,'' \emph{Science}, vol. 333, no. 6051, pp.
  1878--1881, 2011.

\bibitem{ferrara2011crawling}
S.~Catanese, P.~D. Meo, E.~Ferrara, , G.~Fiumara, and A.~Provetti, ``Crawling
  {F}acebook for social network analysis purposes,'' in \emph{Proc.
  International Conf. on Web Intelligence, Mining and Semantics}, 2011.

\bibitem{gjoka2011practical}
M.~Gjoka, M.~Kurant, C.~Butts, and A.~Markopoulou, ``Practical recommendations
  on crawling online social networks,'' \emph{IEEE Journal on Selected Areas in
  Communications}, vol.~29, no.~9, pp. 1872--1892, 2011.

\bibitem{handcock2010modeling}
M.~Handcock and K.~Gile, ``Modeling social networks from sampled data,''
  \emph{The Annals of Applied Statistics}, vol.~4, no.~1, pp. 5--25, 2010.

\bibitem{morstatter2013sample}
F.~Morstatter, J.~Pfeffer, H.~Liu, and K.~Carley, ``{Is the sample good enough?
  Comparing data from Twitters streaming API with Twitters firehose},'' in
  \emph{Proc. Int. Conf. on Weblogs and Social Media}, 2013.

\bibitem{shalizi2011homophily}
C.~Shalizi and A.~Thomas, ``Homophily and contagion are generically confounded
  in observational social network studies,'' \emph{Sociological Methods \&
  Research}, vol.~40, no.~2, pp. 211--239, 2011.

\bibitem{danescu2013no}
C.~Danescu-Niculescu-Mizil, R.~West, D.~Jurafsky, J.~Leskovec, and C.~Potts,
  ``No country for old members: user lifecycle and linguistic change in online
  communities,'' in \emph{Proc. International Conference on World Wide
  Web}.\hskip 1em plus 0.5em minus 0.4em\relax ACM, 2013, pp. 307--318.

\bibitem{schifanella2010folks}
R.~Schifanella, A.~Barrat, C.~Cattuto, B.~Markines, and F.~Menczer, ``Folks in
  folksonomies: social link prediction from shared metadata,'' in \emph{Proc.
  Int. Conf. on Web Search and Data Mining}, 2010, pp. 271--280.

\bibitem{kazienko2010multi}
P.~Kazienko, P.~Brodka, K.~Musial, and J.~Gaworecki, ``Multi-layered social
  network creation based on bibliographic data,'' in \emph{Proc. IEEE
  International Conference on Social Computing}, 2010, pp. 407--412.

\bibitem{tang2012inferring}
J.~Tang, T.~Lou, and J.~Kleinberg, ``Inferring social ties across heterogenous
  networks,'' in \emph{Proc. International Conference on Web search and Data
  Mining}.\hskip 1em plus 0.5em minus 0.4em\relax ACM, 2012, pp. 743--752.

\bibitem{berlingerio2013multidimensional}
M.~Berlingerio, M.~Coscia, F.~Giannotti, A.~Monreale, and D.~Pedreschi,
  ``Multidimensional networks: foundations of structural analysis,''
  \emph{World Wide Web}, vol.~16, no. 5-6, pp. 567--593, 2013.

\bibitem{wang2011learning}
C.~Wang, R.~Raina, D.~Fong, D.~Zhou, J.~Han, and G.~Badros, ``Learning
  relevance from heterogeneous social network and its application in online
  targeting,'' in \emph{Proc. International Conference on Research and
  development in Information Retrieval}.\hskip 1em plus 0.5em minus 0.4em\relax
  ACM, 2011, pp. 655--664.

\bibitem{diehl2007relationship}
C.~Diehl, G.~Namata, and L.~Getoor, ``Relationship identification for social
  network discovery,'' in \emph{Proc. International Conference on Artificial
  Intelligence}, vol.~22, no.~1, 2007, pp. 546--552.

\bibitem{wang2010mining}
C.~Wang, J.~Han, Y.~Jia, J.~Tang, D.~Zhang, Y.~Yu, and J.~Guo, ``Mining
  advisor-advisee relationships from research publication networks,'' in
  \emph{Proc. International Conference on Knowledge Discovery and Data
  Mining}.\hskip 1em plus 0.5em minus 0.4em\relax ACM, 2010, pp. 203--212.

\bibitem{chen2008combinational}
W.~Chen, D.~Zhang, and E.~Chang, ``{Combinational Collaborative Filtering for
  Personalized Community Recommendation},'' in \emph{Proc. International
  Conference on Knowledge Discovery and Data mining}.\hskip 1em plus 0.5em
  minus 0.4em\relax ACM, 2008, pp. 115--123.

\bibitem{chen2009collaborative}
W.~Chen, J.~Chu, J.~Luan, H.~Bai, Y.~Wang, and E.~Chang, ``{Collaborative
  Filtering for Orkut Communities: Discovery of User Latent Behavior},'' in
  \emph{Proc. International Conference on World Wide Web}.\hskip 1em plus 0.5em
  minus 0.4em\relax ACM, 2009, pp. 681--690.

\bibitem{BlNgJo03}
D.~Blei, A.~Ng, and M.~Jordan, ``{Latent Dirichlet Allocation},'' \emph{Journal
  of Machine Learning Research}, vol.~3, pp. 993--1022, 2003.

\bibitem{katz1953new}
L.~Katz, ``A new status index derived from sociometric analysis,''
  \emph{Psychometrika}, vol.~18, no.~1, pp. 39--43, 1953.

\bibitem{ChZeYu13}
L.~Chen, W.~Zeng, and Q.~Yuan, ``A unified framework for recommending items,
  groups and friends in social media environment via mutual resource fusion,''
  \emph{Expert Systems with Applications}, vol.~40, no.~8, pp. 2889--2903,
  2013.

\bibitem{de2010query}
P.~De~Meo, G.~Quattrone, and D.~Ursino, ``A query expansion and user profile
  enrichment approach to improve the performance of recommender systems
  operating on a folksonomy,'' \emph{User Modeling and User-Adapted
  Interaction}, vol.~20, no.~1, pp. 41--86, 2010.

\bibitem{lux2007aspects}
M.~Lux, M.~Granitzer, and R.~Kern, ``Aspects of broad folksonomies,'' in
  \emph{Proc. International Database and Expert Systems Applications}.\hskip
  1em plus 0.5em minus 0.4em\relax IEEE, 2007, pp. 283--287.

\bibitem{heymann2008can}
P.~Heymann, G.~Koutrika, and H.~Garcia-Molina, ``{Can social bookmarking
  improve Web search?}'' in \emph{Proc. International Conference on Web search
  and Web Data Mining}.\hskip 1em plus 0.5em minus 0.4em\relax ACM Press, 2008,
  pp. 195--206.

\bibitem{Cattuto*07}
C.~Cattuto, C.~Schmitz, A.~Baldassarri, V.~Servedio, V.~Loreto, A.~Hotho,
  M.~Grahl, and G.~Stumme, ``Network properties of folksonomies,''
  \emph{Artificial Intelligence Communications}, vol. 20:4, pp. 245--262, 2007.

\bibitem{halpin2007complex}
H.~Halpin, V.~Robu, and H.~Shepherd, ``The complex dynamics of collaborative
  tagging,'' in \emph{Proc. International Conference on World Wide Web}.\hskip
  1em plus 0.5em minus 0.4em\relax ACM Press, 2007, pp. 211--220.

\bibitem{backstrom2006group}
L.~Backstrom, D.~Huttenlocher, J.~Kleinberg, and X.~Lan, ``Group formation in
  large social networks: membership, growth, and evolution,'' in \emph{Proc.
  International Conference on Knowledge Discovery and Data mining}.\hskip 1em
  plus 0.5em minus 0.4em\relax ACM, 2006, pp. 44--54.

\bibitem{jin2011topic}
Y.~Jin, R.~Li, K.~Wen, X.~Gu, and F.~Xiao, ``Topic-based ranking in folksonomy
  via probabilistic model,'' \emph{Artificial Intelligence Review}, vol.~36,
  no.~2, pp. 139--151, 2011.

\bibitem{SiSi09}
S.~Siersdorfer and S.~Sizov, ``{Social recommender systems for Web 2.0
  folksonomies},'' in \emph{Proc. 20th Conference on Hypertext and
  Hypermedia}.\hskip 1em plus 0.5em minus 0.4em\relax ACM, 2009, pp. 261--270.

\bibitem{HoBaBl10}
M.~Hoffman, F.~Bach, and D.~Blei, ``{Online learning for Latent Dirichlet
  Allocation},'' in \emph{Proc. International Conference on Advances in Neural
  Information Processing Systems}, 2010, pp. 856--864.

\bibitem{good2005permutation}
P.~Good, \emph{Permutation, parametric and bootstrap tests of
  hypotheses}.\hskip 1em plus 0.5em minus 0.4em\relax Springer, 2005.

\bibitem{shannon48}
C.~Shannon, ``A mathematical theory of communication,'' \emph{Bell System
  Technical Journal}, vol.~27, 1948.

\bibitem{cover1991elements}
T.~Cover and A.~Thomas, \emph{Elements of information theory}.\hskip 1em plus
  0.5em minus 0.4em\relax Wiley \& Sons, 2012.

\bibitem{scott2009multivariate}
D.~Scott, \emph{Multivariate density estimation: theory, practice, and
  visualization}.\hskip 1em plus 0.5em minus 0.4em\relax Wiley \& Sons, 2009,
  vol. 383.

\end{thebibliography}

\end{document}